\documentclass[reprint,aps,prb,nofootinbib,superscriptaddress]{revtex4-2}

\usepackage[T1]{fontenc}
\usepackage{textcomp}
\usepackage{amsmath,amssymb}

\usepackage{newtxtext,newtxmath}
\usepackage{bm}
\usepackage{graphicx}
\usepackage{braket}
\usepackage{hyperref}
\usepackage{cleveref}
\usepackage{multirow}
\usepackage{algorithm}
\usepackage{algorithmic}
\usepackage{placeins}
\usepackage{makecell}
\usepackage{footmisc}
\usepackage[dvipsnames]{xcolor}

\begin{document}

\title{A Study of Gate-Based and Boson Sampling Quantum Random Number Generation on IBM and Xanadu Quantum Devices}
\author{Mohamed Messaoud \surname{Louamri}}
\thanks{Source code and data available at \url{https://github.com/mmlongh/icpafse2025-qrng-with-ibm-and-xanadu-quantum-devices}}
\affiliation{Constantine Quantum Technologies, University Constantine 1 Frères Mentouri, Constantine, Algeria}
\affiliation{Theoretical Physics Laboratory, USTHB, Algiers, Algeria}

\author{Achraf \surname{Boussahi}}
\affiliation{Constantine Quantum Technologies, University Constantine 1 Frères Mentouri, Constantine, Algeria}
\affiliation{École Supérieure en Informatique 08 Mai 1945, Sidi Bel Abbes, Algeria}

\author{Nacer Eddine \surname{Belaloui}}
\affiliation{Constantine Quantum Technologies, University Constantine 1 Frères Mentouri, Constantine, Algeria}
\affiliation{Laboratoire de Physique Mathématique et Subatomique, University Constantine 1 Frères Mentouri, Constantine, Algeria}

\author{Abdellah \surname{Tounsi}}
\affiliation{Constantine Quantum Technologies, University Constantine 1 Frères Mentouri, Constantine, Algeria}
\affiliation{Laboratoire de Physique Mathématique et Subatomique, University Constantine 1 Frères Mentouri, Constantine, Algeria}

\author{Mohamed Taha \surname{Rouabah}}
\email[Corresponding author: ]{taha.rouabah@cqtech.org}
\affiliation{Constantine Quantum Technologies, University Constantine 1 Frères Mentouri, Constantine, Algeria}
\affiliation{Laboratoire de Physique Mathématique et Subatomique, University Constantine 1 Frères Mentouri, Constantine, Algeria}

\begin{abstract}
Quantum mechanics offers a fundamentally unpredictable entropy source due to the intrinsic probabilistic nature of quantum measurements, making it attractive for secure random number generation. This paper explores the practicality of generating random numbers from two quantum platforms: gate-based circuits on IBM Quantum and (Gaussian) boson sampling with Xanadu Borealis. We implement simple post-processing methods, including the classic Von Neumann extractor and two tailored variants designed to address the correlated structure of boson sampling outputs. We evaluate debiased output from real quantum hardware using the NIST SP800-22r1a test suite and measure the extraction efficiency of each debiasing method. Results show that, while unbiased bitstreams can be achieved on both platforms, throughput remains low and cost per random bit is high compared to specialized QRNG devices.
\end{abstract}
\maketitle

\section{Introduction}

Random numbers are essential in applications such as cryptography, stochastic simulations, and scientific computing. Despite their apparent simplicity, generating truly random numbers---both unpredictable and unbiased---at high rates is challenging~\cite{Dale2015}. Unpredictability means future values cannot be inferred, even with full system knowledge, while unbiasedness requires all possible outcomes to be equally likely.

Classical pseudo-random number generators (e.g., linear congruential generators~\cite{knuth1981seminumerical}) produce reproducible sequences from deterministic rules, which fundamentally violates the unpredictability constraint for cryptographic applications. Hardware random number generators attempt to solve this issue by sampling from physical entropy sources (e.g., thermal noise, radioactive decay, etc.). However, without post-processing, raw hardware randomness can produce biased results that compromise randomness quality~\cite{harvard}. Post-processors fix this by eliminating systematic skews and restoring uniform probability across all possible outputs.

Quantum mechanics provides a fundamentally unpredictable source of randomness through measurement outcomes on superposition states. This provides a fundamental mechanism for random number generation, known as quantum random number generation (QRNG), which can be implemented either through dedicated devices\footnote{For an example of commercial QRNG devices, see ID Quantique: \url{https://www.idquantique.com/random-number-generation/overview/}.} or by repurposing general-purpose quantum systems.

Our work evaluates the practicality of generating random numbers from two quantum platforms: gate-based quantum computers via IBM Quantum~\cite{IBMQuantum}, and Gaussian boson samplers from Xanadu~\cite{XanaduCloud}. Then, it introduces post-processing methods designed to handle their respective bias and correlation structures.

The paper is organized as follows: \cref{sec:qrng-gbqc} and \cref{sec:qrng-bs} discuss QRNG using gate-based and boson sampling quantum devices, respectively; \cref{sec:exp-dem} presents results using IBM's Sherbrooke and Xanadu's Borealis quantum devices; and \cref{sec:conclusion} presents the work's conclusion.

\section{QRNG Using Gate-Based Quantum Computers} \label{sec:qrng-gbqc}

Gate-based quantum computers use qubits, which can exist in superpositions of \(\ket{0}\) and \(\ket{1}\). Initialized to a known state (typically \(\ket{0}\)), qubits are manipulated by quantum gates to create probabilistic measurement outcomes. For example, applying the Hadamard gate, $H$, where
\begin{equation}
    H = \frac{1}{\sqrt{2}}\begin{bmatrix} 1 & 1 \\ 1 & -1 \end{bmatrix}, \quad H\ket{0} = \frac{1}{\sqrt{2}}(\ket{0} + \ket{1}),
\end{equation}
produces a state that gives an equal probability of obtaining \(0\) or \(1\) when measured, thus yielding an unbiased random bit in the ideal case. In practice, however, quantum devices are noisy. Decoherence, relaxation, and readout errors will skew the output distribution, making it necessary to debias the results.

\subsubsection*{Debiasing with Von Neumann's Method:}

A classic technique for removing bias is the Von Neumann (VN) algorithm~\cite{von_1951_various}, which processes bitstreams in pairs:
\begin{equation}
    00 \text{ or } 11 \rightarrow \text{discarded};\quad 01 \rightarrow 0;\quad 10 \rightarrow 1.
\end{equation}
For example, the sequence \(S=00\,11\,10\,10\,01\) yields the output 110.

Assuming input bits are \textbf{independent} with $\Pr(0)=p$, we can prove that this technique gives unbiased outputs and that the probability of retaining a pair is $p(1-p)$, giving an \emph{extraction efficiency} (ExE)
\begin{equation}
    \text{ExE} \equiv \frac{\text{len(output)}}{\text{len(input)}} \approx p(1-p) \leq 0.25.
\end{equation}

This bit yield is quite low. Improved methods, such as Elias’s generalization~\cite{peter} and block-based schemes with waiting strategies~\cite{zonga}, can raise efficiency to over 60\% while preserving uniformity. 

\section{QRNG Using Boson Samplers}\label{sec:qrng-bs}
Boson Sampling and Gaussian Boson Sampling are quantum systems where photonic input states---either single photons or squeezed states---are sent through an \( M \)-mode linear optical network composed of beam splitters and phase shifters~\cite{Aaronson2011}. Measurement at the output yields photon counts in each mode. In our case, we are only interested in the presence (\(1\)) or absence (\(0\)) of photons, not their count.

Unlike the previous section’s independent bitstreams, these outputs are correlated within each measurement due to photon number conservation and mode entanglement. This violates the VN method’s bit-independence assumption, so applying VN as-is would bias the output.

To address this problem, we propose two variants of post-processing that adapt the VN logic to work with boson sampling data.

\subsubsection*{Variant 1 -- Single Mode Comparison}
We select a single mode (e.g., the first column), and apply VN to bits across successive shots (\cref{algo:variant_1}):
\begin{algorithm}[H]
\caption{Variant 1  -- Single Mode Comparison}
\begin{algorithmic}[1]
\REQUIRE Binary matrix \texttt{bits} of shape (shots, 1, modes)
\STATE Initialize empty list \texttt{unbiased\_bits}
\FOR{each pair of successive shots $(i, i+1)$}
    \STATE $b_1 \gets \texttt{bits}[i, 0, 0]$
    \STATE $b_2 \gets \texttt{bits}[i+1, 0, 0]$
    \IF{$b_1 \neq b_2$}
        \STATE Append $b_1$ to \texttt{unbiased\_bits}
    \ENDIF
\ENDFOR
\RETURN \texttt{unbiased\_bits}
\end{algorithmic}
\label{algo:variant_1}
\end{algorithm}

This avoids inter-mode correlations but discards most of the data, resulting in low efficiency.

\subsubsection*{Variant 2 -- Multi Mode Comparison}
We compare multiple output modes from two different shots, selecting the first bit pair that differs:
\begin{algorithm}[H]
\caption{Variant 2 -- Multi Mode Comparison}
\begin{algorithmic}[1]
\REQUIRE Binary matrix \texttt{bits} of shape (shots, 1, modes)
\STATE Initialize empty list \texttt{unbiased\_bits}
\FOR{each pair of successive shots $(i, i+1)$}
    \STATE $s_1 \gets \texttt{bits}[i, 0, :]$
    \STATE $s_2 \gets \texttt{bits}[i+1, 0, :]$
    \FOR{each $(b_1, b_2)$ in $s_1, s_2$}
        \IF{$b_1 \neq b_2$}
            \STATE Append $b_1$ to \texttt{unbiased\_bits}
            \STATE \textbf{break}
        \ENDIF
    \ENDFOR
\ENDFOR
\RETURN \texttt{unbiased\_bits}
\end{algorithmic}
\end{algorithm}

This method increases efficiency while maintaining unbiasedness. Both variants provide provably unbiased bitstreams tailored to the structure of boson sampling data, with Variant 2 offering significantly better ExE, as can be seen in \cref{tab:bit_extraction_summary}.

\section{Experimental Demonstrations} \label{sec:exp-dem}
We now move to experimental validation, where we generate or reuse quantum hardware data, apply suitable post-processing, and test the resulting output's randomness using the NIST SP800-22r1a suite.~\cite{nistsuite}.

\subsection{IBM Sherbrooke}
Using the 127-qubit IBM Sherbrooke device, we applied Hadamard gates to all qubits and sampled the circuit 100,000 times, generating 12.7 million raw bits (50.46\% zeros). The raw bitstring failed the Monobit test ($p = 2.75 \times 10^{-236}$), prompting the use of Von Neumann debiasing. This yielded 3,169,704 unbiased bits (24.96\% efficiency). The total quantum execution time was approximately 35 seconds. At a pay-as-you-go rate of \$96 per minute on IBM Quantum~\cite{IBMQuantum}, this corresponds to a cost of \$56. The effective generation rate was about 90{,}563 unbiased bits per second, at a cost of roughly \$17.67 \textit{per million unbiased bits}. While the output passes the statistical tests (see \cref{tab:nist_all_pvalues}), these numbers highlight the prohibitive cost and limited scalability of gate-based quantum hardware for high-throughput random number generation\footnote{For reference, the ID~Quantique Quantis QRNG PCIe 40\,Mbps device costs around~\$1,655. \textbf{This is the cost of the hardware itself---not individual randomness samples}. So, the effective cost per million bits is very low.}.

\subsection{Xanadu Borealis}
Although Borealis has been sunsetted, we repurposed data from Xanadu's quantum computational advantage experiments~\cite{Madsen2022}. We applied the two proposed post-processing methods to extract unbiased bits. \Cref{tab:bit_extraction_summary} summarizes the datasets, shot counts, and extraction yields. Assuming a pay-as-you-go cost of \$100 per million shots, the 84,096,000-shot dataset from Fig.~2\footnote{All figure references point to~\cite{Madsen2022}; this work contains no figures.} corresponds to a total execution cost of approximately \$8,410. With Variant 2 post-processing, gives cost of roughly \$200 \textit{per million unbiased bits}. This is significantly more expensive than dedicated QRNG hardware.

\onecolumngrid

\begin{table}[htbp]
\centering

\begin{tabular}{lcccc}
\hline
\textbf{Dataset} & \textbf{Shots} & \textbf{Postprocessing Variant} & \textbf{Bits} & \textbf{ExE (\%)} \\
\hline
\multirow{2}{*}{Fig. 2 (16-mode)} & \multirow{2}{*}{84,096,000} & Variant 1 & 16,685,995 & 19.84\% \\
                                  &                              & Variant 2 & 42,007,537 & 49.95\% \\
\hline
\multirow{2}{*}{Fig. 3a (216-mode)} & \multirow{2}{*}{1,044,000} & Variant 1 & 91,451 & 8.76\% \\
                                    &                            & Variant 2 & 522,000 & 50.00\% \\
\hline
\multirow{2}{*}{Fig. 3b (72-mode)} & \multirow{2}{*}{1,036,000} & Variant 1 & 202,026 & 19.50\% \\
                                   &                            & Variant 2 & 518,000 & 50.00\% \\
\hline
\multirow{2}{*}{Fig. 4 (216-mode)} & \multirow{2}{*}{1,026,000} & Variant 1 & 246,580 & 24.03\% \\
                                   &                            & Variant 2 & 513,000 & 50.00\% \\
\hline
\multirow{2}{*}{Fig. S15 (288-mode)} & \multirow{2}{*}{1,113,000} & Variant 1 & 256,764 & 23.07\% \\
                                     &                            & Variant 2 & 556,500 & 50.00\% \\
\hline
\end{tabular}
\caption{Extraction efficiency from Xanadu Borealis datasets using two post-processing methods.}
\label{tab:bit_extraction_summary}
\end{table}

\begin{table}[H]
\centering
\begin{tabular}{lccc}
\hline
\textbf{Test} & \textbf{IBM Sherbrooke} & \textbf{Borealis (Variant 1)} & \textbf{Borealis (Variant 2)} \\
\hline
Monobit Test & 0.8122 & 0.3292 & 0.0313 \\
\hline 
Approximate Entropy Test & 0.5728 & 0.7448 & 0.5480 \\
\hline 
Binary Matrix Rank Test & 0.7675 & 0.6012 & 0.9972 \\
\hline 
Cumulative Sums Test & 0.5673 & 0.0522 & 0.0152 \\
\hline 
Discrete Fourier Transform Test & 0.1443 & 0.6265 & 0.1966 \\
\hline 
Frequency Within Block Test & 0.5080 & 0.4327 & 0.6797 \\
\hline 
Longest Run of Ones in a Block Test & 0.9102 & 0.2104 & 0.2056 \\
\hline 
Overlapping Template Matching Test & 0.6964 & 0.4527 & 0.1014 \\
\hline 
Random Excursion Test & 0.0949 & 0.1131 & 0.2366 \\
\hline 
Random Excursion Variant Test & 0.0329 & 0.1924 & 0.0650 \\
\hline 
Runs Test & 0.2743 & 0.6740 & 0.7554 \\
\hline 
Serial Test & 0.4955 & 0.7386 & 0.5483 \\
\hline 
Maurer’s Universal Test & 0.3277 & 0.2680 & 0.1315 \\
\hline 
\makecell[l]{Non-Overlapping Template\\Matching Test} & 1.0000 & 1.0000 & 1.0000 \\
\hline 
Linear Complexity Test & 0.9237 & 0.4816 & 0.0422 \\
\hline
\end{tabular}
\caption{NIST SP800-22r1a p-values for 2M-bit substrings from IBM Sherbrooke and Xanadu Borealis (two post-processing variants) results. \textbf{All tests were passed}.}
\label{tab:nist_all_pvalues}
\end{table}
\twocolumngrid
\subsection{Randomness Testing}
We selected a 2 million-bit substring  from the beginning of each  post-processed stream and evaluated it using the NIST SP800-22r1a suite. All tests were passed in every case, as summarized in \cref{tab:nist_all_pvalues}.

\FloatBarrier
\section{Conclusion} \label{sec:conclusion}
This work assessed the feasibility of using contemporary quantum hardware as sources for secure random number generation. By combining standard and tailored post-processing methods, we demonstrated that both gate-based and boson-sampling quantum devices can yield statistically-sound randomness, provided appropriate debiasing techniques are applied. However, our results underscore a significant tradeoff: while randomness quality is high, throughput rate and cost-effectiveness remain limiting factors for large-scale deployment. In particular, general-purpose quantum processors are far less economical than dedicated QRNG hardware, both in cost-per-bit and execution latency.

Nonetheless, our findings provide a practical reference for the capabilities and limitations of current devices in randomness extraction tasks. The proposed extraction variants offer simple strategies for handling correlated outputs.

\section*{Acknowledgements}
This document has been produced with the financial assistance of the European Union (Grant no. DCI-PANAF/2020/420-028), through the African Research Initiative for Scientific Excellence (ARISE), pilot programme. ARISE is implemented by the African Academy of Sciences with support from the European Commission and the African Union Commission. The contents of this document are the sole responsibility of the author(s) and can under no circumstances be regarded as reflecting the position of the European Union, the African Academy of Sciences, and the African Union Commission.

We are grateful to the Algerian Ministry of Higher Education and Scientific Research and DGRST for the financial support.

We gratefully acknowledge the use of IBM Quantum systems through public free-tier access available at the time. We also acknowledge Xanadu’s openly published~\cite{Madsen2022} Gaussian Boson Sampling datasets.

\end{document}